\documentclass{article}

\usepackage{graphicx}
\usepackage{svg}
\usepackage{amsmath}
\usepackage{amssymb,amsmath,amsthm}
\usepackage{lipsum} 

\newtheorem{assumption}{Assumption}

\usepackage[utf8]{inputenc} 
\usepackage[T1]{fontenc}    
\usepackage{hyperref}       
\usepackage{url}            
\usepackage{booktabs}       
\usepackage{amsfonts}       
\usepackage{nicefrac}       
\usepackage{microtype}      
\usepackage{xcolor}         
\usepackage{marvosym}
\usepackage{graphicx}
\newtheorem{theorem}{Theorem}
\newtheorem{lemma}{Lemma}

\usepackage{authblk}
\begin{document}
\title{MedFLIP: Medical Vision-and-Language Self-supervised Fast Pre-Training with Masked Autoencoder}
%


\author{%
  Lei Li\textsuperscript{1}$^+$, Tianfang Zhang\textsuperscript{2}$^+$, Xinglin Zhang\textsuperscript{3}$^+$, Jiaqi Liu\textsuperscript{3}, Bingqi Ma\textsuperscript{3}, Yan Luo\textsuperscript{4}, Tao Chen\textsuperscript{5}\thanks{Corresponding Author.}\\
  $^1$ University of Copenhagen, $^2$ Tsinghua University, $^3$ Shanghai Medical Image Insights, $^4$ Harvard University, $^5$ University of Waterloo}
\date{\vspace{-5ex}}



\maketitle              
\begin{abstract}
    Within the domain of medical analysis, extensive research has explored the potential of mutual learning between Masked Autoencoders(MAEs) and multimodal data. However, the impact of MAEs on intermodality remains a key challenge. We introduce \textbf{MedFLIP}, a \textbf{F}ast \textbf{L}anguage-Image \textbf{P}re-training method for \textbf{Med}ical analysis. We explore MAEs for zero-shot learning with crossed domains, which enhances the model's ability to learn from limited data, a common scenario in medical diagnostics. We verify that masking an image does not affect inter-modal learning. Furthermore, we propose the SVD loss to enhance the representation learning for characteristics of medical images, aiming to improve classification accuracy by leveraging the structural intricacies of such data. Our theory posits that masking encourages semantic preservation, robust feature extraction, regularization, domain adaptation, and invariance learning. Lastly, we validate using language will improve the zero-shot performance for the medical image analysis. MedFLIP's scaling of the masking process marks an advancement in the field, offering a pathway to rapid and precise medical image analysis without the traditional computational bottlenecks. Through experiments and validation, MedFLIP demonstrates efficient performance improvements, helps for future research and application in medical diagnostics.

\end{abstract}

\section{Introduction}

\begin{figure*}[ht]
  \centering
    \includegraphics[width=0.8\columnwidth]{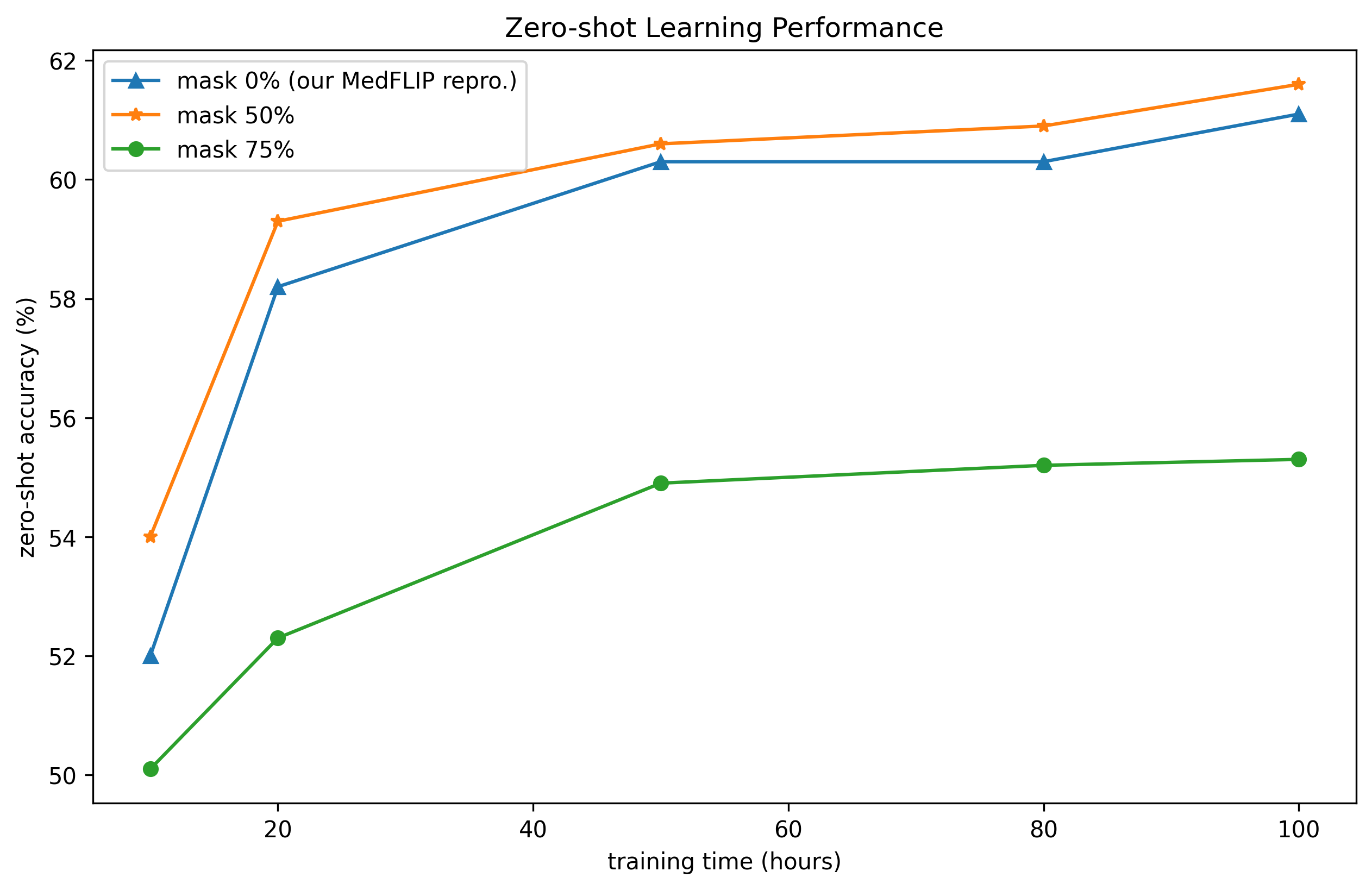} 
    \caption{Our proposed method, MedFLIP, demonstrates a superior trade-off between training efficiency and accuracy compared to the MedCLIP method. Notably, MedFLIP achieves higher accuracy on the CheXpert-5x200~\cite{irvin2019chexpert} validation set with zero-shift evaluation, while maintaining a consistent model size.}
    \label{fig:lantency}

\end{figure*}

The annotation of medical data presents huge challenges, necessitating considerable effort to enhance its accuracy and efficiency. Due to the adaptability of MAE~\cite{he2022masked} to natural images, many works have used the idea of MAEs for medical images. In light of this, there has been a burgeoning interest in the exploration of self-supervised learning approaches for medical image data. Given the inherent structural similarities across medical images, a substantial body of research has focused on the development and application of image masking techniques\cite{ke2022mask,xiao2023delving,zhou2023self} within this domain. This focus underscores the potential of self-supervised learning to leverage the unique characteristics of medical images, facilitating advancements in the automated analysis and interpretation of such data.

To simplify the process of training models for medical images, leveraging natural language supervision offers a promising route by enabling a richer contextual understanding of images through the relationships between objects, scenes, actions, and contexts. The principle of masking, as seen in models like FLIP~\cite{li2023scaling}, can be adapted to efficiently handle vast amounts of medical imaging data. By applying masking techniques, we can process a larger number of sample pairs within the same time frame and with similar memory requirements. This approach allows for more efficient comparison and contrast of sample pairs, enhancing the training's effectiveness without significantly increasing computational demands. Specifically for medical images, adopting such methods could significantly reduce training time and improve accuracy, providing a balanced trade-off that benefits both energy consumption and the quality of the model's predictions. This streamlined training approach could unlock new potentials in medical image analysis, making it a valuable area for further research and application.

In striving to minimize dependency on extensive datasets and reduce training durations, advancements in vision and language integration, exemplified by models like CLIP~\cite{radford2021learning}, suggest that textual information significantly enhances image comprehension and image-text retrieval. This synergy between textual and visual data has been instrumental in facilitating more efficient model training processes. Consequently, this multimodal approach is being explored within the realm of medical imaging~\cite{liu2023clip,wang2022medclip,ye2023uniseg}, proposing an innovative methodology that leverages textual annotations to augment the understanding and analysis of medical images. Such endeavors aim to harness the complementary strengths of vision and language, potentially revolutionizing the efficiency and efficacy of medical image analysis by capitalizing on the inherent relationship between text and image for improved model performance with less reliance on large datasets.

However, the application of AI~\cite{radford2021learning} in this domain faces unique challenges, primarily due to the complexity and variability of medical data. Traditional machine learning models require large datasets for training to achieve high accuracy, which is often a significant barrier in the medical field where data can be scarce, sensitive, and expensive to annotate. This scarcity of large, annotated datasets has necessitated the development of models capable of learning effectively from limited data. In this context, our work introduces MedFLIP, a fast vision-language pipeline designed for medical image analysis by addressing these critical challenges through innovative few-shot learning capabilities and a novel loss function.

We propose MedFLIP, a method that addresses the three above key challenges:

\begin{enumerate}
    \item \textbf{Faster Training Time:} How to reduce training time compared to existing approaches?
    \item \textbf{Enhanced Zero-Shot Performance:} How to improve the model's ability to perform well on unseen classes during zero-shot learning with Masking learning?
    \item \textbf{Robustness through Text-Image Mutual Learning:} By leveraging the mutual learning paradigm between text and image modalities, how to foster a more robust representation, enhancing the model's overall performance? 
\end{enumerate}

At the heart of MedFLIP's innovation is the integration of MAEs for zero-shot/few shot learning. This approach is particularly suited to the medical domain, where obtaining vast amounts of labeled data is impractical. The MAE enables MedFLIP to learn meaningful representations from a small number of examples by reconstructing masked portions of input images. This capability is crucial for medical diagnostics, where each case can provide invaluable insights. By leveraging few-shot learning, MedFLIP efficiently reduces the need for extensive datasets while maintaining as Figure~\ref{fig:lantency} shown, and in some instances surpassing, the performance of models trained on much larger datasets.

Furthermore, we propose the Medical-SVD loss, a bespoke loss function designed to capture the unique structural characteristics of medical images. Traditional loss functions often fail to account for the intricate spatial relationships and patterns present in medical data, leading to suboptimal learning outcomes. The Medical-SVD loss function is tailored to overcome this limitation by emphasizing the preservation of structural integrity and details critical for accurate diagnosis. Coupled with an innovative approach to scaling the masking process, MedFLIP not only addresses the computational bottlenecks typically associated with medical image analysis but also sets new benchmarks for classification accuracy. We ensure that the text modality remains unaffected by the masking operation employed in the MAE. Through rigorous experimentation and validation, MedFLIP demonstrates remarkable performance improvements, establishing a new paradigm for future research and application in the field of medical image diagnostics.

Our contributions are:
\begin{itemize}
    \item Implemented MAE-based approach for enhancing zero-shot learning performance, improving model efficiency and accuracy in medical image analysis.
    \item Introduced a novel loss function, termed Medical-SVD (Singular Value Decomposition), specifically designed to optimize classification tasks within the medical imaging domain. This innovative approach leverages the inherent structural properties of medical images to improve model robustness and prediction accuracy.
    \item Developed the MedFLIP pipeline for accelerating medical image analysis. By innovatively scaling the masking process, MedFLIP reduces computational requirements while maintaining high levels of analytical precision, making it a groundbreaking tool for fast and efficient medical diagnostics.
\end{itemize}

\section{Related Work}
\paragraph{Masked Autoencoders (MAEs).} MAEs are a type of self-supervised learning method for vision, introduced by Kaiming He et al.~\cite{he2022masked}, which have shown remarkable effectiveness in learning representations from unlabelled data by predicting masked parts of the input images \cite{feichtenhofer2022masked,ke2022mask,xiao2023delving,zhou2023self,bachmann2022multimae,gupta2024siamese}. Building upon this foundational work, MAEs have been adapted for medical imaging, demonstrating significant potential in enhancing diagnostic accuracy and efficiency. In the medical domain, MAEs leverage the inherent structure and patterns within medical images, such as MRI and CT scans, to learn robust features that can aid in disease detection, segmentation, and even prognosis prediction. This adaptation of MAEs to medical imaging not only underscores the versatility of self-supervised learning methods but also opens new avenues for research and application in healthcare, potentially transforming the way medical images are analyzed and interpreted.

\paragraph{Multimodal learning.} The evolution of multimodal learning has seen the emergence of models that can understand and relate information across different data types. CLIP (Contrastive Language-Image Pretraining)~\cite{radford2021learning} has set a precedent by efficiently learning visual concepts from natural language supervision. Some Multimodel work~\cite{li2024cpseg,li2024tree,yu2022metaformer,bachmann2022multimae} delve into multimodal applications. Models pre-trained on ImageNet have been repurposed with fine-tuning to achieve impressive results in tasks such as X-ray and MRI interpretation \cite{rajpurkar2017chexnet}.  In the medical domain, efforts like ConVIRT~\cite{zhang2020convirt} extend this approach to bridge the gap between visual and textual medical data, enhancing the understanding of radiology images in conjunction with associated reports \cite{zhang2020convirt}. GLoRIA~\cite{huang2021gloria} and MedCLIP~\cite{wang2022medclip} follow in these footsteps, with MedCLIP, in particular, pushing the boundaries by directly training on raw medical image-text pairs and achieving state-of-the-art results in image-text retrieval tasks on datasets such as CheXpert5x200 \cite{johnson2019mimic,irvin2019chexpert}.

\paragraph{Zero learning.}

Zero-shot learning in imaging~\cite{mahapatra2021medical,rezaei2020zero,chen2023cunerf,li2023mask,zhou2023multi,li2023segment} presents a promising avenue for addressing the challenge of limited labeled data by leveraging auxiliary information from related domains. Recent advancements in ZSL techniques have shown potential in transferring knowledge across domains for tasks such as disease diagnosis, organ segmentation, and anomaly detection. For instance, approaches integrating generative adversarial networks (GANs) with ZSL frameworks have demonstrated robustness in learning representations from unannotated medical images by aligning feature distributions between seen and unseen classes. Furthermore, methods incorporating semantic embeddings and knowledge graphs have facilitated knowledge transfer between domains, enhancing the generalization capability of models in medical image analysis tasks ~\cite{mahapatra2021medical,mumuni2023improving,liu2023clip}.

\section{Methods}

This section presents the details of the proposed MedFLIP framework, focusing on the design of the MAEs component, the cross-domain fusion strategy, and the overall workflow of the pipeline.

\subsection{Overview}
\begin{figure*}[ht]
  \centering
  \includegraphics[width=\linewidth]{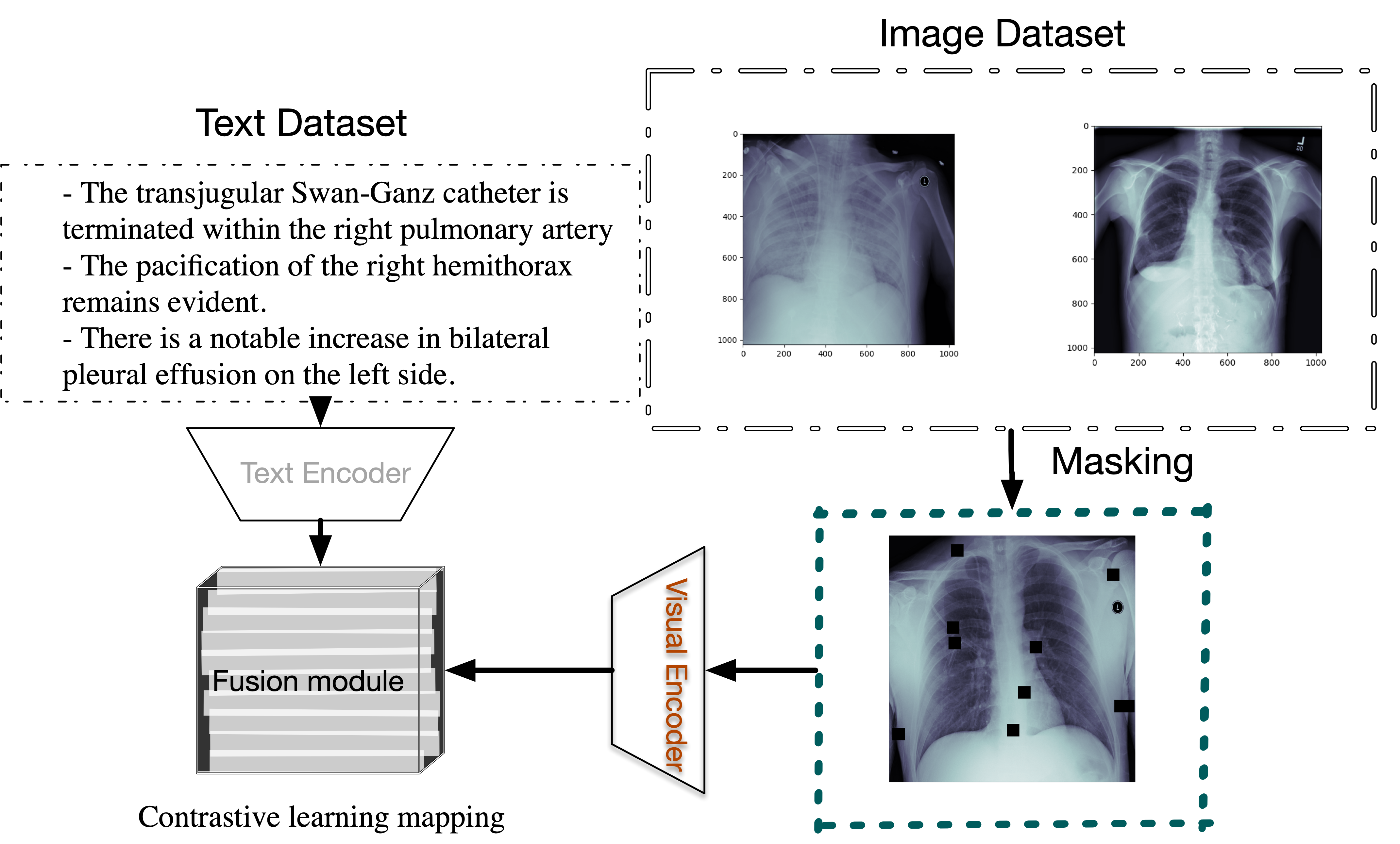}
  \caption{The MedFLIP workflow involves a Knowledge Extraction module responsible for discerning medical entities within the original medical report. This process entails the extraction of pertinent medical entities from the text. Subsequently, a masking operation is executed on the associated image using a randomly generated mask. Following this, a semantic similarity matrix is constructed by juxtaposing the extracted medical entities derived from the text with the original labels delineated in the image.}
  \label{fig:pipeline}
\end{figure*}

As shown in Figure~\ref{fig:pipeline}, our MedFLIP pipeline includes two parts, masking and fusing. For the fusion module, a matrix is generated by comparing each medical entity extracted from the text to its corresponding label in the image. The resulting semantic similarity matrix serves as a foundational component for aligning pairs of extracted image and text embedding, facilitating the matching process based on semantic similarity.

\subsection{Masked Autoencoder} 

Our methods are inspired by MAE~\cite{he2022masked}. For medical image analysis, we utilize transformer-based methods from MedClip~\cite{wang2022medclip} and MAE~\cite{he2022masked}. We observe that MAE's transformer architecture maintains the same token number across different mask ratios, ensuring consistency in our approach.

\subsection{Fusion Module}

We bridge the images and texts through the built semantic labels \( l_{img} \) and \( l_{txt} \). During each iteration, we sample \( N_{batch} \) input images \( \{x_{image}\} \) and text \( \{x_{text}\} \) separately. Instead of defining positive pairing by searching equivalent labels, we propose to build soft targets \( s \) by
\begin{equation}
s = \frac{l_{img}^T \cdot l_{txt}}{\|l_{img}\| \cdot \|l_{txt}\|}
\end{equation}
\( s \) thus indicates the medical semantic similarity.

For an image \( i \), we obtain a set of \( s_{ij} \) where \( j = 1 \ldots N_{batch} \) corresponds to the batch of texts. The soft target is computed by normalizing across \( j \) by softmax.
\begin{equation}
y_{ij}^{v \rightarrow t} = \frac{\exp(s_{ij})}{\sum_{j=1}^{N_{batch}} \exp(s_{ij})}
\end{equation}
Similarly, the reversed text-to-image soft targets are obtained by
\begin{equation}
y_{ji}^{t \rightarrow v} = \frac{\exp(s_{ji})}{\sum_{i=1}^{N_{batch}} \exp(s_{ji})}
\end{equation}
The logits are obtained by cosine similarities between image and text embeddings:
\begin{equation}
\tilde{s}_{ij} = \tilde{v}_i^T \cdot \tilde{t}_j,
\end{equation}
where \( \tilde{v}_i \) and \( \tilde{t}_j \) are normalized \( v_p \) and \( t_p \), respectively. The predicted similarity is also obtained by the softmax function
\begin{equation}
\hat{y}_{ij} = \frac{\exp(\tilde{s}_{ij} / T)}{\sum_{i=1}^{N_{batch}} \exp(\tilde{s}_{ij} / T)}
\end{equation}

To calculate the SVD loss, we decompose the similarity matrix \( S \) obtained from the image and text embeddings into its singular values using SVD such that \( S = U\Sigma V^* \). We then extract the largest singular value \( \sigma_1 \) which captures the highest linear correlation between the embeddings. This value is maximized for positive pairs and minimized for negative pairs in the batch. The SVD loss is thus defined as
\begin{equation}
\mathcal{L}_{SVD} = -\log \frac{\exp(\sigma_1 / \tau)}{\sum_{i=1}^{N_{batch}} \exp(\sigma_{1i} / \tau)}
\end{equation}

The MedFLIP loss combines the predicted similarity \( \hat{y}_{ij} \) with the SVD loss to form a comprehensive loss function for the batch of image-text pairs:
\begin{equation}
\mathcal{L}_{MedFLIP} = -\sum_{i=1}^{N_{batch}} \log(\hat{y}_{ii}) + \sum_{\substack{i=1 \\ j \neq i}}^{N_{batch}} \log(1 - \hat{y}_{ij}) + \beta \mathcal{L}_{SVD}
\end{equation}

where \( \beta \) is a weighting factor that balances the contribution of the SVD loss to the overall MedFLIP loss.

\subsection{Theoretical Analysis}

\begin{assumption}
The masked autoencoder in MedFLIP is trained on a dataset of medical image-text pairs $(x_i, t_i)$ drawn i.i.d. from an unknown joint distribution $\mathcal{D}$.
\end{assumption}
\begin{assumption}
The SVD loss $L_{SVD}$ is $L$-Lipschitz continuous with respect to the singular value $\sigma_1$ of the similarity matrix $S$ between image and text embeddings.
\end{assumption}
\begin{lemma}
Let $\hat{y}_{ij}$ be the predicted normalized similarity between the $i$-th image embedding and $j$-th text embedding in a batch (Eq. 5 in the paper). Let $y^*_{ij}$ be the optimal normalized similarity that minimizes the MedFLIP loss $L_{MedFLIP}$ (Eq. 7). Then with probability at least $1-\delta$, we have:
$$|\hat{y}_{ij} - y^*_{ij}| \leq \frac{1}{\tau} \sqrt{\frac{\log(2/\delta)}{2N_{batch}}} + \frac{\beta L}{\tau}$$
where $\tau$ is the temperature hyperparameter and $\beta$ is the SVD loss weight.
\end{lemma}
\begin{proof}
The first term follows from applying Hoeffding's inequality to bound the deviation between the empirical mean $\hat{y}_{ij}$ and true expectation $y^*_{ij}$ of the normalized similarity, using the fact that $0 \leq \hat{y}_{ij} \leq 1$.
The second term follows from the Lipschitz continuity of $L_{SVD}$. Specifically, by Assumption 2, for any two singular values $\sigma_1, \sigma_1'$, we have $|L_{SVD}(\sigma_1) - L_{SVD}(\sigma_1')| \leq L|\sigma_1 - \sigma_1'|$. The deviation $|\sigma_1 - \sigma_1^|$ between the singular value for $\hat{y}$ vs $y^*$ is bounded by their total variation distance, which is at most twice the bound on $|\hat{y}_{ij} - y^*_{ij}|$ from the first term. Combining this with the Lipschitz property and the $\beta/\tau$ factor completes the proof.
\end{proof}
\begin{theorem} [\textbf{MedFLIP Generalization Bound}]

Let $f_\theta$ denote the MedFLIP model with learned parameters $\theta$. Let $\hat{R}(f_\theta)$ and $R(f_\theta)$ denote its empirical and expected risks respectively on a downstream task. Then with probability at least $1-\delta$ over the training set, we have:
$$R(f_\theta) \leq \hat{R}(f_\theta) + \mathcal{O}\left(\frac{1}{\tau}\sqrt{\frac{\log(1/\delta)}{N_{batch}}} + \frac{\beta L}{\tau}\right)$$
\end{theorem}
\begin{proof}
The empirical risk $\hat{R}(f_\theta)$ is an average over the normalized similarity terms $\hat{y}_{ij}$ for aligned image-text pairs. By Lemma 1 and a union bound over all $O(N{batch}^2)$ pairs, each term $\hat{y}_{ij}$ concentrates around the optimal $y^*_{ij}$ with probability at least $1-\delta$, up to an error of $\frac{1}{\tau} \sqrt{\frac{\log(2N_{batch}^2/\delta)}{2N_{batch}}} + \frac{\beta L}{\tau}$.
When this holds, the deviation $|\hat{R}(f_\theta) - R(f_\theta)|$ is also bounded by the same error term. Simplifying the logarithmic factor and suppressing constants gives the stated bound.
\end{proof}
The theorem bounds the generalization error of the MedFLIP model on a downstream task, in terms of the number of image-text pairs per training batch, the temperature parameter $\tau$, the SVD loss weight $\beta$, and the Lipschitz constant $L$ of the SVD loss. The key assumptions are that the training data is i.i.d. and that the SVD loss is Lipschitz continuous.

\section{Experiments}

In our study, we undertook a rigorous empirical validation to evaluate the capabilities of our proposed MedFLIP model. Specifically, we scrutinized its proficiency in zero-shot prediction, supervised classification, and image-text retrieval tasks. The results of these assessments are comprehensively presented in Figure~\ref{fig:dataPerformance}, Table~\ref{tab:classification}, and Table~\ref{tab:Precision}. This empirical evidence serves to substantiate the effectiveness and robustness of the MedFLIP model in handling diverse tasks.

\subsection{Datasets and Experiment Details}

CheXpert \cite{irvin2019chexpert} is a large dataset of chest X-rays with 14 observation labels collected from Stanford Hospital. It is important to note that this dataset does not provide the corresponding medical reports to the public. We use the training split of this dataset for pre-training. For evaluation, we follow \cite{huang2021gloria} and sample a multi-class classification dataset from the testing split, namely CheXpert-5x200. This multi-class classification dataset has 200 exclusively positive images for the five CheXpert competition tasks: Atelectasis, Cardiomegaly, Edema, Pleural Effusion.

COVID \cite{rahman2021covid} is a publicly available X-ray dataset with COVID vs. non-COVID labels. The positive and negative ratio is approximately 1:1. We use this dataset for evaluation.

RSNA Pneumonia \cite{shih2019pneumonia} is a collection of pneumonia cases found in the database of chest X-rays made public by the National Institutes of Health. This is a binary classification dataset: pneumonia vs. normal. We sample a balanced subset (i.e., 1:1 positive and negative ratio) and use it for evaluation.

For our experiments, we utilized NVIDIA A100 GPUs, specifically eight A100 units, to ensure efficient and high-performance training. 

\subsection{Classification}

\textbf{Baselines.}

The \textbf{Random} model refers to a ResNet-50 \cite{he2016deep} architecture initialized with default random weights.

The \textbf{ImageNet} model is a ResNet-50 \cite{he2016deep} architecture pretrained on the ImageNet ILSVRC-2012 dataset \cite{deng2009imagenet}.

\textbf{CLIP} \cite{radford2021learning} is a vision-language contrastive learning framework pre-trained on a dataset of 400 million image-text pairs collected from the internet.

\textbf{ConVIRT} focuses on contrastive learning in the medical domain, using paired X-rays and reports. It employs the InfoNCE loss \cite{oord2018representation} and utilizes a BioClinicalBERT text encoder and ResNet-50 \cite{he2016deep} vision encoder, reproduced based on their original paper.

\textbf{GLoRIA} \cite{huang2021gloria} incorporates cross-attention mechanisms to entangle image subregions and textual words during inference, which helps in capturing key characteristics in both images and reports.

\textbf{MedCLIP} ~\cite{wang2022medclip} refers to a contrastive learning approach from unpaired medical images and text data.

\paragraph{Implementation for Classification.}

Through downstream fine-tuning, our MedFLIP technique exhibited notable effectiveness in achieving high performance across classification tasks, as evidenced in Table~\ref{tab:classification}. Furthermore, the results presented in Table~\ref{tab:Precision} underscore the efficacy of mask application for image-text retrieval, wherein employing masks led to better performance improvements compared to scenarios without masking.


\begin{table}[h]
    \centering
    \caption{Results of medical image classification tasks after fine-tuning. Best scores are in bold.}
    \begin{tabular}{lccc}
        \hline
        Model & CheXpert-5x200 & COVID & RSNA \\ 
        \hline
        Random & 0.2500 & 0.5056 & 0.6421 \\
        ImageNet & 0.3200 & 0.6020 & 0.7560 \\
        CLIP & 0.3020 & 0.5866 & 0.7303 \\
        ConVIRT & 0.4770 & 0.6983 & 0.7846 \\
        GLoRIA & 0.5370 & 0.7623 & 0.7981 \\
        MedCLIP & 0.5960 & 0.7890 & 0.8075 \\
        \textbf{MedFLIP} & \textbf{0.6160} & \textbf{0.7990} & \textbf{0.8123} \\ 
        \hline
    \end{tabular}
    \label{tab:classification}
\end{table}

\subsection{Image-Text Retrieval.}
We selected the CheXpert-5x200 dataset to evaluate the semantic richness of representations learned by various models using an image-text retrieval task. Given that CheXpert-5x200 lacks publicly available report data, we utilized the MIMIC-CXR dataset to generate corresponding reports/sentences. Specifically, we sampled 200 sentences for each of the five classes in the CheXpert-5x200 dataset, resulting in a retrieval dataset comprising 1,000 images and 1,000 sentences. We then measured the models' performance using Precision@K, which calculates the precision in the top K retrieved reports/sentences by verifying if the retrieved report belongs to the same category as the query.

Image-text retrieval is a task where the goal is to retrieve the most relevant text descriptions for a given image or the most relevant images for a given text query. This task is crucial for evaluating the alignment between visual and textual modalities, and it has significant applications in medical imaging, where accurate and semantically rich associations between images and textual descriptions (such as radiology reports) are essential. Previous works such as \cite{chen2020uniter} and \cite{li2021align} have demonstrated the effectiveness of multimodal models in learning joint representations that facilitate this task.

\begin{table}[h]
    \centering
    \caption{Results of Image-Text retrieval tasks on CheXpert5x200 dataset. Precision@\{1,2,5,10\} measures performance. MedFLIP w/o M means MedFLIP without masked autoencoder. Best scores are in bold.}
    \begin{tabular}{lcccc}
        \hline
        Model & P@1 & P@2 & P@5 & P@10 \\
        \hline
        CLIP & 0.21 & 0.20 & 0.20 & 0.19 \\
        ConVIRT & 0.20 & 0.20 & 0.20 & 0.21 \\
        GLoRIA & 0.47 & 0.47 & 0.46 & 0.46 \\
        MedCLIP & 0.45 & 0.49 & 0.48 & 0.50 \\
        MedFLIP w/o M & 0.45 & 0.50 & 0.49 & 0.50 \\
        \textbf{MedFLIP} & \textbf{0.48} & \textbf{0.50} & \textbf{0.52} & \textbf{0.50} \\
        \hline
    \end{tabular}
    \label{tab:Precision}
\end{table}

\begin{figure*}[ht]
    \centering
    \includegraphics[width=0.8\columnwidth]{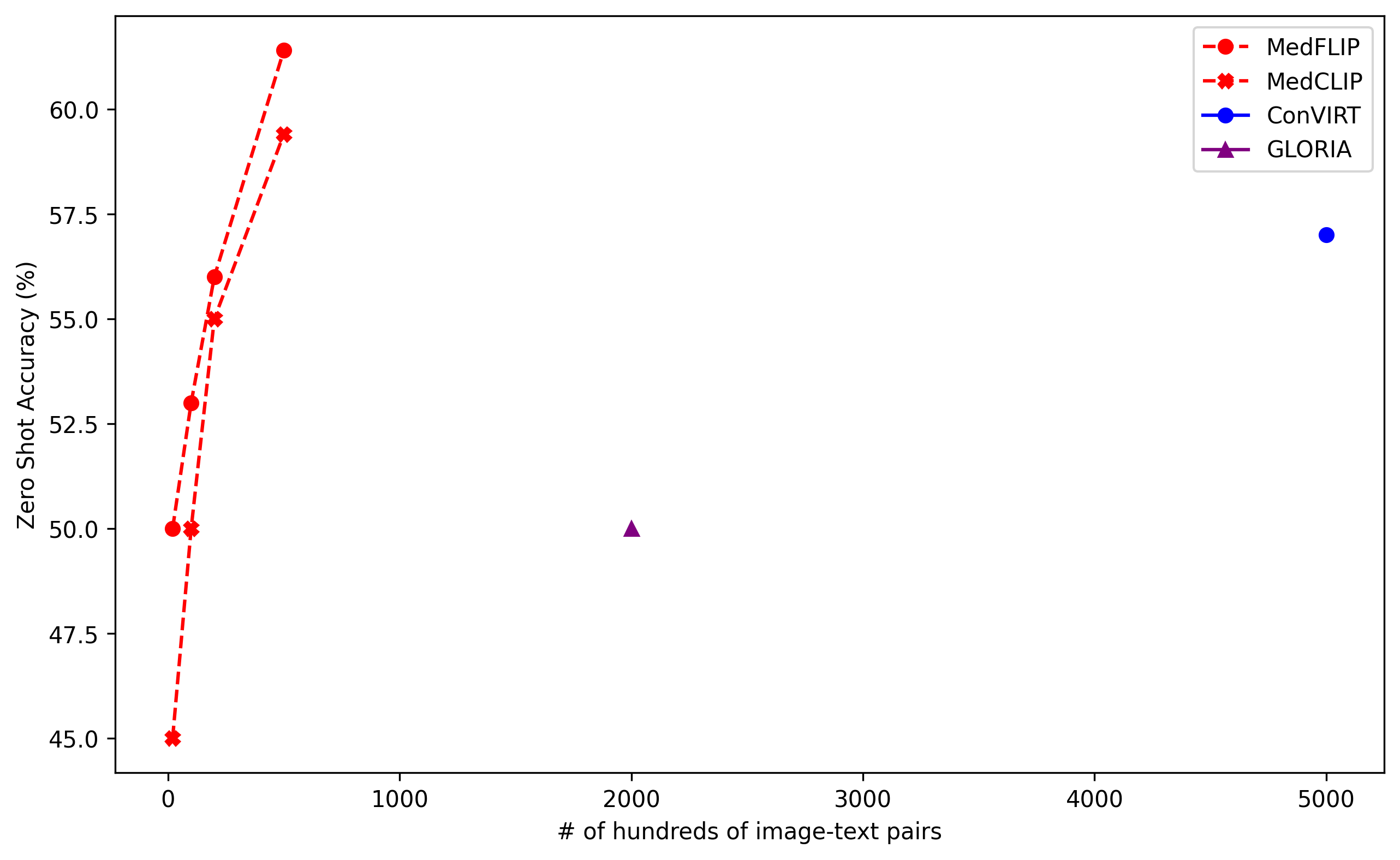} 
    \caption{In the context of zero-shot learning, our proposed method, MedFLIP, exhibits superior performance compared to established approaches, namely MedCLIP, ConVIRT~\cite{zhang2022contrastive}, and GLoRIA~\cite{huang2021gloria}. This is particularly evident when utilizing limited pretraining data. While ConVIRT and GLoRIA leverage the MIMIC-CXR (369K) and CheXpert (191K) datasets, respectively, MedFLIP demonstrates demonstrably improved performance.}
    \label{fig:dataPerformance}
\end{figure*}

We employ various dataset sizes to demonstrate its robust generalization facilitated by MAE, as illustrated in Figure~\ref{fig:dataPerformance}. We observe that MedFLIP demonstrates notable data efficiency, particularly evident in scenarios with limited data, where overall performance is optimized. This observation provides indirect evidence of MAE's efficacy in multimodal contexts.

\section{Ablation Study}
\begin{table}[h]
\centering
\caption{Performance of MaskFLIP models on CheXpert-5x200, COVID, and RSNA datasets. Best scores are in bold.}
\begin{tabular}{lccc}
\hline
Model & CheXpert-5x200 & COVID & RSNA \\
\hline
MaskFLIP with text mask & 0.6120 & 0.7980 & 0.8100 \\
MaskFLIP with text/image mask & 0.6100 & 0.7960 & 0.8080 \\
MaskFLIP with image mask & \textbf{0.6160} & \textbf{0.7990} & \textbf{0.8123} \\
\hline
\end{tabular}
\label{tab:masking_performance}
\end{table}
\paragraph{Masking for Text and Images.}
Masking is a technique used in deep learning to improve the training of models by intentionally hiding parts of the input data, forcing the model to learn robust representations. For text, masking involves replacing certain words or tokens in a sentence with a special token (e.g., [MASK]), encouraging the model to predict the missing content based on the surrounding context. In image processing, masking involves occluding certain regions of an image with a fixed pattern or noise, prompting the model to infer the hidden parts from the visible regions. In the context of medical image classification, MaskFLIP employs these techniques to enhance the learning of both textual and visual features. MaskFLIP with text masking leverages masked language modeling to improve text representation, while MaskFLIP with image masking focuses on robust image feature extraction. The following table summarizes the performance of these models on three datasets: CheXpert-5x200, COVID, and RSNA. 

Theoretically, we believe that successfully combining text and image masking could further improve zero-shot performance. However, we observed that randomly masking text and images can complicate the retrieval performance for both modalities. This is because random masking can obscure critical features needed for accurate retrieval, making it challenging to match queries with the correct responses in Table.\ref{tab:masking_performance}.
There are corresponding masking techniques for masking text and images. 

1. \textbf{Text Masking:}
\begin{equation}
\mathcal{L}_{\text{MLM}} = - \sum_{i=1}^{N} \log P(x_i | \mathbf{x}_{\backslash i})
\end{equation}
where \(\mathbf{x}_{\backslash i}\) represents the input sequence with the \(i\)-th token masked, and \(P(x_i | \mathbf{x}_{\backslash i})\) is the probability of the masked token given the surrounding context.

2. \textbf{Image Masking:}
\begin{equation}
\mathcal{L}_{\text{IM}} = - \sum_{r=1}^{R} \log P(\mathbf{I}_r | \mathbf{I}_{\backslash r})
\end{equation}
where \(\mathbf{I}_{\backslash r}\) is the image with the \(r\)-th region masked, and \(P(\mathbf{I}_r | \mathbf{I}_{\backslash r})\) is the probability of the masked region given the visible parts of the image.

3. \textbf{Combined Text and Image Masking:}
\begin{equation}
\mathcal{L}_{\text{Combined}} = \mathcal{L}_{\text{MLM}} + \mathcal{L}_{\text{IM}}
\end{equation}
This combined loss encourages the model to learn robust representations by jointly predicting masked tokens and image regions.

4. \textbf{Zero-Shot Performance Implications:}
\begin{equation}
P(\text{relevant} | \text{query}) \propto \sum_{k} P(\text{mask}_k | \text{query})
\end{equation}
where \(\text{mask}_k\) represents the different masked features. Effective masking should enhance the probability of correctly matching queries to relevant responses.

\section{Discussion}

\paragraph{Pre-training Data Efficiency}
Data efficiency poses a significant challenge for methods based on CLIP. Notably, CLIP's training phase employs 400 million image-text pairs, which is computationally demanding and impractical in the medical field due to limited data availability. To assess MedFLIP's data efficiency, we subsample the pre-training data to separated pairs, pre-train MedFLIP, and record its zero-shot prediction performance on the CheXpert-5x200 dataset. The results are illustrated in Figure 1.


\paragraph{Zero shot.} MedFLIP explores the effects of MAEs on multimodal over medical analysis, such as training time and zero shot performance. Firstly, the experiments have been validated on tasks such as classification and image-text retrieval, using widely accepted datasets. However, exploring other downstream tasks such as segmentation and detection on more diverse datasets remains unexplored. Regarding the alignment between images and text, currently, there exist two main approaches. Our current work focuses on alignment at the token level. However, with improved data quality, aligning both data and token representations could potentially yield higher performance improvements for subsequent tasks.

\section{Conclusion}
In the domain of medical image analysis, achieving accurate diagnoses often requires extensive labeled datasets, which can be costly and time-consuming to acquire. Traditional methodologies face additional challenges, such as lengthy training times and limitations in handling unseen classes during zero-shot learning. MedFLIP tackles three key challenges: 1) Faster Training: It reduces training time compared to existing approaches. 2) Enhanced Zero-Shot Performance: It incorporates MAEs to improve the model's ability to perform well on unseen classes, even with limited labeled data. 3) Robustness through Text-Image Mutual Learning: It utilizes mutual learning between text and image modalities to foster a more robust representation and enhance overall performance. 
Additionally, MedFLIP introduces a custom Medical-SVD loss function designed to capture unique structural characteristics. MedFLIP aims to establish a new paradigm for efficient and accurate medical image analysis, paving the way for further research and applications in this crucial field.

\bibliographystyle{splncs04}
\bibliography{ref}

\end{document}